\documentclass[journal=apchd5,manuscript=article]{achemso}
\setkeys{acs}{articletitle = true,doi = true}
\setkeys{acs}{
    etalmode = truncate,
    maxauthors = 10
}

\usepackage[version=3]{mhchem}
\usepackage{booktabs}
\usepackage{bm}
\usepackage{lmodern,microtype}
\usepackage[scaled=0.875]{helvet}
\usepackage[dvipsnames]{xcolor}
\usepackage[bookmarks=false,colorlinks]{hyperref}
\hypersetup{linkcolor=magenta,citecolor=MidnightBlue,filecolor=Plum,urlcolor=MidnightBlue}
\usepackage{subcaption}
\usepackage{tikz}

\newcommand{\supt}{\textcolor{RoyalBlue}{Table}\,S}

\def\checkmark{\tikz\fill[scale=0.4](0,.35) -- (.25,0) -- (1,.7) -- (.25,.15) -- cycle;}

\author{Aubrey G.\ J.\ Nyiri}
\affiliation[Northwestern University]{Department of Materials Science and Engineering, Northwestern University, Evanston, Illinois 60208, USA}

\author{Michael J.\ Waters}
\affiliation[Northwestern University]{Department of Materials Science and Engineering, Northwestern University, Evanston, Illinois 60208, USA}
\email{michael.j.waters@northwestern.edu}

\author{James M.\ Rondinelli}
\affiliation[Northwestern University]{Department of Materials Science and Engineering, Northwestern University, Evanston, Illinois 60208, USA}
\email{jrondinelli@northwestern.edu}

\title[Upper bound paper]
  {A Normalized Descriptor for Unbiased Screening of Second-Order Nonlinear Optical Materials}

\abbreviations{NLO, SHG, ML, PBE, LDA, HSE, LBO, NIR, UV, DUV}
\keywords{nonlinear optics, second-harmonic generation, normalized descriptor, ab initio, high-throughput screening, machine learning}

\begin{document}

\begin{abstract}
Second-order nonlinear optical materials enable frequency doubling of light (second-harmonic generation, SHG), which is essential for optoelectronic applications ranging from materials characterization to quantum technologies. 
However, comparing SHG performance across materials remains challenging as the second-order nonlinear susceptibility $\chi^{(2)}$ spans several orders of magnitude and strongly depends on the band gap $E_g$. 
To address this, we empirically validate a theoretical upper bound on $\chi^{(2)}$ using new databases of \textit{ab initio}-computed nonlinear optical (NLO) properties.
We then formulate a normalized descriptor, $\hat{d}$, which expresses the NLO response of a material relative to the band gap-dependent physical limit. 
We show that $\hat{d}$ exhibits a similar distribution across a wide range of band gap energies.
This universality supports the use of $\hat{d}$ as a robust, generalizable descriptor for data-driven and chemistry-informed machine learning models of NLO response, enabling accelerated materials discovery and optimization across broad application frequencies.  
\end{abstract}

\section{Introduction}
Nonlinear optical (NLO) materials are valued for their frequency conversion capabilities, leading to advancements in the fields of microscopy, information storage, quantum computing, and more\cite{garmire2013, Aslam2023}. 
The second-order nonlinear response, responsible for second-harmonic generation (SHG), is particularly important. 
However, maximizing the performance of NLO materials for SHG is plagued by conflicting relationships between desired properties: second-order susceptibility tensor $\chi^{(2)}_{{ijk}}$ (or in Voigt contracted matrix  notation $d_{{ij}}=\frac{1}{2}\chi^{(2)}_{{ijk}}$), high laser-induced damage threshold (related to the band gap, $E_g$, and thermal conductivity $\kappa$), moderate birefringence $\Delta n$, optical transparency across the application wavelength range, and mechanical stability \cite{boyd1989, Halasyamani2018, Aslam2023}.
A primary consideration is that $\chi^{(2)}$ typically decreases with increasing $E_g$, often spanning 2-3 orders of magnitude across the range of band gaps. 
This also complicates direct comparisons of the SHG response between materials with different band gaps, when elucidating materials features, properties, and structures that influence performance \cite{Rondinelli2015}. 
This is particularly important to address for machine learning (ML)-based screening approaches \cite{Petousis2017,Naccarato2019,trinquet2025,Hassan2025}, where a single scalar performance label is helpful.
Theoretical studies have established relationships between the second-order nonlinear susceptibility $\chi^{(2)}$ and the band gap energy $E_g$, offering a physically grounded framework for scaling the second-harmonic generation (SHG) response \cite{taghizadeh2021, kuzyk2000, kuzyk2006, lytel2017, tan2019, dawson2015}. 
Notably, Taghizadeh, Thygesen, and Pederson (TTP) derived an expression for the upper bound on the diagonal tensor components ($i=j=k$) of the static second-order nonlinear susceptibility tensor for crystalline compounds using a two-band model as\cite{taghizadeh2021}
\begin{equation}
    |\chi^{(2)}_{iii}|\leq\frac{24e^3E_0^2\Xi}{\epsilon_0}\cdot E_g^{-4}.
    \label{eq:chi2_upper_bound}
\end{equation}
Here, $E_0$ determines the magnitude of the hopping amplitude for an electron, \textit{e.g.} within a tight-binding model, and $\Xi$ is a dimensionless geometrical factor determined by the lattice, hopping ranges, and measurement direction \cite{tan2019}. 
Importantly, the scaling of this upper bound differs from the $E_g^{-3.5}$ dependence derived for molecules by Kuzyk \cite{kuzyk2000,Lytel2016}. 
This discrepancy arises from fundamental differences in electronic structure: crystalline materials exhibit continuous band dispersions, whereas molecules possess discrete energy levels.
These distinctions influence the application of the Thomas–Reiche–Kuhn (TRK) sum rule, which underpins the derivation of the upper bound in both cases \cite{taghizadeh2021, mossman2016}.
%
The upper bound on $\chi^{(2)}$ stems from ``the fact that transition dipole matrix elements in a quantum system cannot be arbitrarily large.\cite{taghizadeh2021}''
Recently, several high-throughput \textit{ab initio} databases of NLO materials for SHG have been developed\cite{xie2023, trinquet2025, yu2020, wang2024}, sparking the use of ML and statistical methods to accelerate the discovery of improved NLO materials. 
Prior studies have employed metrics such as $d_{\mathrm{KP}}$, the rotational average of the SHG tensor measured via the Kurtz-Perry (KP) powder technique\cite{kurtz1968, Clark2022, alkabakibi2025} and $d_{ij}^{\mathrm{max}}$, the maximum component of the SHG tensor \cite{ran2025}, as target values for predictive modeling. 
Screening based on these metrics, however, may be overly simplistic.
The inverse relationship between $\chi^{(2)}$ and $E_g$ inherently biases the search against promising materials with larger band gaps. 
Accordingly, as database sizes grow in the field of nonlinear optics, it is vital to properly pursue the multiobjective nature of the system to select screening criteria in an unbiased manner.\cite{Low2023}
To address this, we first empirically validate the theoretical scaling relationship for the TTP upper bound of $\chi^{(2)}$ using new \textit{ab initio} data for pristine three-dimensional to zero-dimensional crystals. 
Building on this foundation, we introduce a normalized SHG descriptor that quantifies a material’s SHG response relative to its band gap–dependent theoretical maximum.
We then demonstrate how this descriptor enables insightful comparison of SHG performance across a wide range of band gaps and serves as an interpretable label for ML-based screening of NLO materials.

\section{Methods}


Band gap and $\chi^{(2)}$ data were compiled from three distinct sources, each employing different exchange-correlation functionals ($V_{xc}$) within density functional theory (DFT), as summarized in \autoref{tab:datasets}:
(1) Trinquet \latin{et al.} ($\sim2{,}200$ entries)\cite{trinquet2025}, 
(2) Yu \latin{et al.} ($\sim1{,}000$ entries) \cite{yu2020}, and 
(3) Wang \latin{et al.} ($\sim2{,}400$ entries).\cite{wang2024}
All datasets report DFT-level calculations of $\chi^{(2)}$, but differ in their choice of functional: Trinquet \latin{et al.} employed the local density approximation (LDA),\cite{trinquet2025} while Yu \latin{et al.} and Wang \latin{et al.} used the generalized gradient approximation (GGA) in the form of the Perdew–Burke–Ernzerhof (PBE) functional.\cite{yu2020, wang2024}

\begin{table}[b]
    \centering
        \caption{DFT approximations to different optical properties for data compiled from the three data sources used in our combined datasets.}
    \begin{tabular}{l l c c c}
        \toprule
        {Property} & $V_{xc}$ & {Trinquet \latin{et al.}}\cite{trinquet2025} & {Yu \latin{et al.}}\cite{yu2020} & {Wang \latin{et al.}}\cite{wang2024} \\ \midrule
        {$E_g$} & LDA & \checkmark &  &  \\ 
        {$E_g$} & PBE & \checkmark & \checkmark & \checkmark \\ 
        {$E_g$} & HSE & \checkmark &  & \checkmark \\
        \midrule
        {$d_{ij}$} & LDA & \checkmark &  &  \\ 
        {$d_{ij}$} & PBE &  & \checkmark & \checkmark \\ 
        {$d_{ij}$} & Scissor-corrected$^*$ & \checkmark &  & \checkmark \\ 
        \bottomrule
        \multicolumn{4}{c}{\small *scissor shift found using HSE band gap as reference} \\
    \end{tabular}
    \label{tab:datasets}
\end{table}

To integrate the LDA- and PBE-based datasets for subsequent analysis, we first identified 378 overlapping entries between the datasets of Trinquet \latin{et al.} and Wang \latin{et al.} 
A linear regression model was then constructed to estimate $d_{ij}^{\mathrm{max}}$ at the PBE level for the Trinquet \latin{et al.} dataset, yielding an $R^2=0.793$, indicative of reasonably strong correlation. 
For entries present in both datasets, only the values from Wang \latin{et al.} were retained in the merged dataset to ensure consistency.
Notably, $\chi^{(2)}$ values from Yu \latin{et al.} were calculated at a wavelength of 1,064.03 nm\cite{yu2020}, while \autoref{eq:chi2_upper_bound} was derived for the static limit, \textit{i.e.}, zero-frequency/long wavelength limit ($\omega\rightarrow0$/$\lambda\rightarrow \infty$).
To minimize the influence of resonant effects on the magnitude of $\chi^{(2)}$, we only included materials from Yu \latin{et al.} with $E_g>3\ \mathrm{eV}$ ($\sim2.5$ times the energy of a photon of wavelength $1{,}064.03\ \mathrm{nm}$) in the combined dataset, eliminating $\sim 100$ entries. 
Both datasets from Trinquet \latin{et al.} and Wang \latin{et al.} were calculated at the zero frequency limit. 
Included data consists of 0-D through 3-D materials. Note that while data from Trinquet \latin{et al.} and Yu \latin{et al.} includes full $\chi^{(2)}$ tensors, available data from Wang \latin{et al.} only includes the maximum tensor component $\chi^{(2)}_{\mathrm{max}}$.\cite{trinquet2025, yu2020, wang2024} 
Accordingly, all of our analysis will use $\chi^{(2)}_{\mathrm{max}}$ to ensure fair comparison of data.

We used additional datasets from Trinquet \textit{et al.} ($\sim700$ entries) and Wang \textit{et al.} ($\sim200$ entries) with band gaps computed via the HSE hybrid functional\cite{trinquet2025, wang2024}. 
These band gaps were used to define scissor shifts applied to the PBE/LDA ground-state electronic structures for the additional $\chi^{(2)}$ calculations,\cite{Nastos2005} yielding a smaller but higher-fidelity dataset for comparison.
The merged dataset is included in Supporting Information (SI).

\section{Results and Discussion}

\subsection{Theoretical Upper Bound Assessment}

\begin{figure}[t]
    \includegraphics[width=0.975\linewidth]{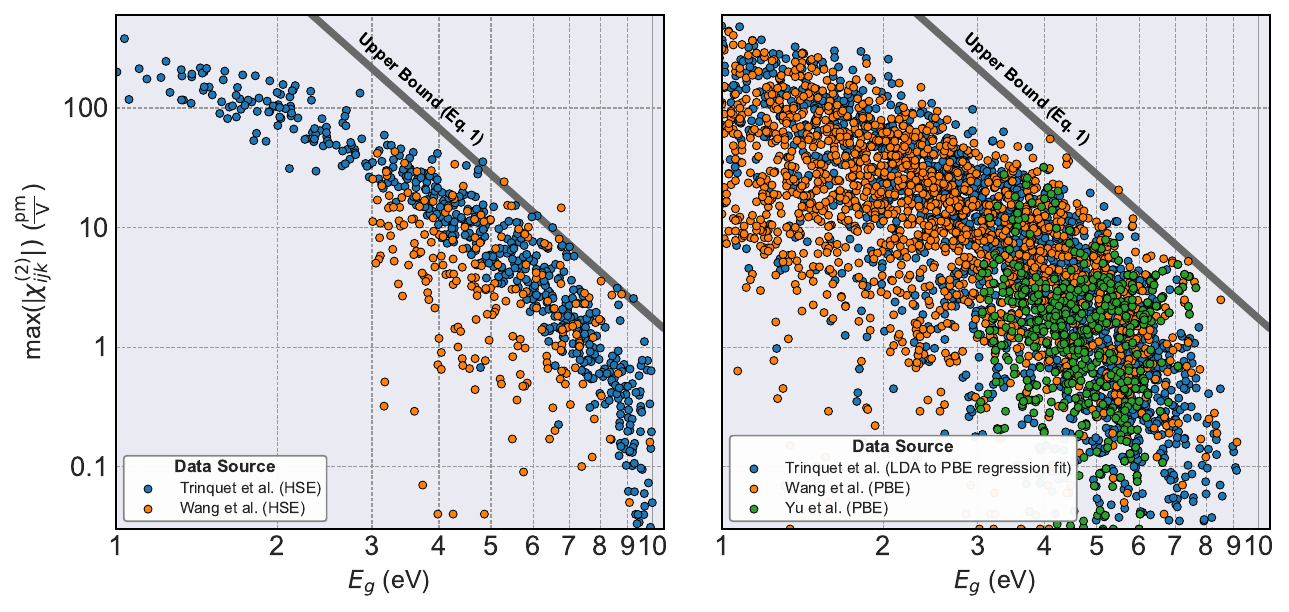}
    \caption{The HSE/scissor-corrected dataset (left) and the PBE dataset (right) plotted against the TTP theoretical upper bound given by \autoref{eq:chi2_upper_bound}. The entries at the Pareto front generally track the shape of the upper bound curve.}
    \label{fig:bound}
\end{figure}

\autoref{fig:bound} presents the HSE/scissor-corrected and PBE datasets alongside the theoretical upper bound for the second-order susceptibility, as defined by \autoref{eq:chi2_upper_bound}. 
The bound is plotted using $\Xi = 1$ and $E_0 = 0.2\ \mathrm{eV}$, which have been previously demonstrated to effectively encompass the majority of known materials.\cite{taghizadeh2021, tan2019}
While these parameter choices capture the general trend and provide a useful reference for comparison, several materials in our dataset exhibit maximum $\chi^{(2)}_{ijk}$ coefficients that exceed the plotted bound. 
This deviation underscores the material-specific nature of $\Xi$ and $E_0$, which are empirically derived and not universally applicable.
Accordingly, the bound in \autoref{eq:chi2_upper_bound} should not be interpreted as a strict limit for all materials. 
Rather, it serves as a benchmark that reflects the underlying scaling behavior, enabling meaningful comparisons across diverse material classes.

Despite some deviations, the theoretical $E_g^{-4}$ scaling behavior predicted by the upper bound in \autoref{eq:chi2_upper_bound} is largely consistent with the empirical data, particularly for materials with band gaps exceeding 3\,eV.
This observation supports the utility of the bound as a normalization framework for the SHG coefficient, enabling the definition of a band gap-independent metric to assess the intrinsic nonlinear optical performance of materials.

\subsection{SHG Performance Descriptor}

We define a normalized SHG coefficient, $\hat{d}$, which expresses the maximum coefficient in the NLO susceptibility of a material relative to the theoretical upper bound based on its band gap $E_g$ as 
\begin{equation}
    \hat{d}(d_{ij}^{\mathrm{max}},\space E_g)=\frac{d_{ij}^{\mathrm{max}}}{\frac{1}{2}\chi^{(2)}_{\mathrm{\mathrm{lim}}}(E_g)}\,,
    \label{eq:d_hat_def}
\end{equation}
where $d_{ij}^{\mathrm{max}}$ is the maximum SHG matrix component and $\chi^{(2)}_{\mathrm{lim}}(E_g)$ is the theoretical maximum second-order nonlinear susceptibility according to \autoref{eq:chi2_upper_bound} (with $\Xi=1$ and $E_0=0.2\ \mathrm{eV}$.)
This results in a simplified version of \autoref{eq:chi2_upper_bound} where $|\chi^{(2)}_{iii}|\leq\xi\cdot E_g^{-4}$ with $\xi=17{,}370\ \mathrm{pm\,V}^{-1}\mathrm{eV^4}$.
Based on \autoref{eq:chi2_upper_bound} and \autoref{eq:d_hat_def}, the final relationship for the scaled SHG coefficient is then given by
\begin{equation}
    \hat{d}(d_{ij}^{\mathrm{max}},\space E_g)=\frac{2}{\xi}\cdot d_{ij}^{\mathrm{max}}\cdot E_g^4\,.
    \label{eq:d_hat_simplified}
\end{equation}
Our formulation effectively maps the NLO response of a material to a dimensionless scale between $0$ and approximately unity, owing to limitations imposed on using a single value of $\Xi$ and $E_0$. 
\autoref{tab:top_ten} lists the ten compounds with the largest values of $\chi^{(2)}_{ijk}$ for the HSE dataset, ranked by the metric $\hat{d}$, each approaching or exceeding the theoretical upper bound.
Shown in \autoref{tab:top_ten}, the experimentally unobserved phase of YOF in the half-Heusler structure (space group ${F\bar{4}3m}$) exceeds the upper bounds by nearly a factor of two.
We hypothesize that this may be attributed to the narrow valence band, shown in the entry for YOF (mp-38194) in the Materials Project database,\cite{horton2025, jain2013} which influences the hopping range $\xi_h$.\cite{tan2019}
Given the strong dependence of the geometrical factor $\Xi$ on $\xi_h$, the unusually large $\hat{d}$ predicted for YOF may be reasonably explained by the influence of its electronic structure on $\Xi$.
The top entries in the PBE dataset are shown in \supt1, and a large dataset of experimental $\hat{d}$ values for complex chalcogenides is available at \url{https://mtd.mccormick.northwestern.edu/IR-NLO-database}.

\begin{table}[t]
    \centering
        \caption{The 10 compounds in the HSE/scissor-corrected dataset with $\chi^{(2)}_{ijk}$ values closest to or exceeding the upper bound, including space group (SG), property values, and relevant notes.}
    \begin{tabular}{l l c c c l}
        \toprule
        \textbf{} & \textbf{SG (\#)} & \textbf{\bm{$E_g$} (eV)} & \textbf{\bm{$d_{ij}^{\mathrm{max}}$} (\bm{$\frac{\mathrm{pm}}{\mathrm{V}}$})} & \bm{$\hat{d}$} & \textbf{Note} \\ \midrule
        \textbf{$\mathrm{YOF}$} & ${F\overline{4}3m}$ (216) & 6.76 & 7.31 & 1.76 & Phase not observed experimentally* \\  
        \textbf{$\mathrm{LiNbO_3}$} & ${R3c}$ (161) & 4.83 & 17.7 & 1.11 & Known NLO material\cite{zhang2020} \\  
        \textbf{$\mathrm{CIN}$} & ${R3m}$ (160) & 5.29 & 12.1 & 1.09 & 1-D structure*, molecular solid \\ 
        \textbf{$\mathrm{SrB_4O_7}$} & ${Pmn2_1}$ (31) & 9.22 & 1.27 & 1.06 & Known NLO material\cite{wu2024} \\ 
        \textbf{$\mathrm{H_4Br_2}$} & ${P1}$ (1) & 6.52 &  5.00 & 1.04 & Molecular solid solution\\ 
        \textbf{$\mathrm{LiHBrCl}$} & ${R3m}$ (160) & 6.10 & 6.27 & 1.00 & No info. available \\
        \textbf{$\mathrm{Ba_2B_8O_{14}}$} & ${Pmn2_1}$ (31) & 8.60 & 1.55 & 0.98 & No info. available \\
        \textbf{$\mathrm{Ba(BH)_{12}}$} & ${P31c}$ (159) & 7.00 & 3.44 & 0.95 & 0-D structure* \\ 
        \textbf{$\mathrm{BPO_4}$} & ${I\overline{4}}$ (82) & 8.76 & 1.39 & 0.94 & Known NLO material\cite{li2016} \\ 
        \textbf{$\mathrm{PNF_2}$} & ${Cmc2_1}$ (36) & 6.70 & 4.01 & 0.93 & 1-D structure* \\ \bottomrule
        \multicolumn{6}{c}{\small *according to Materials Project database\cite{horton2025, jain2013}} \\
    \end{tabular}
    \label{tab:top_ten}
\end{table}

\begin{figure}[t]
\centering
    \includegraphics[width=0.99\linewidth]{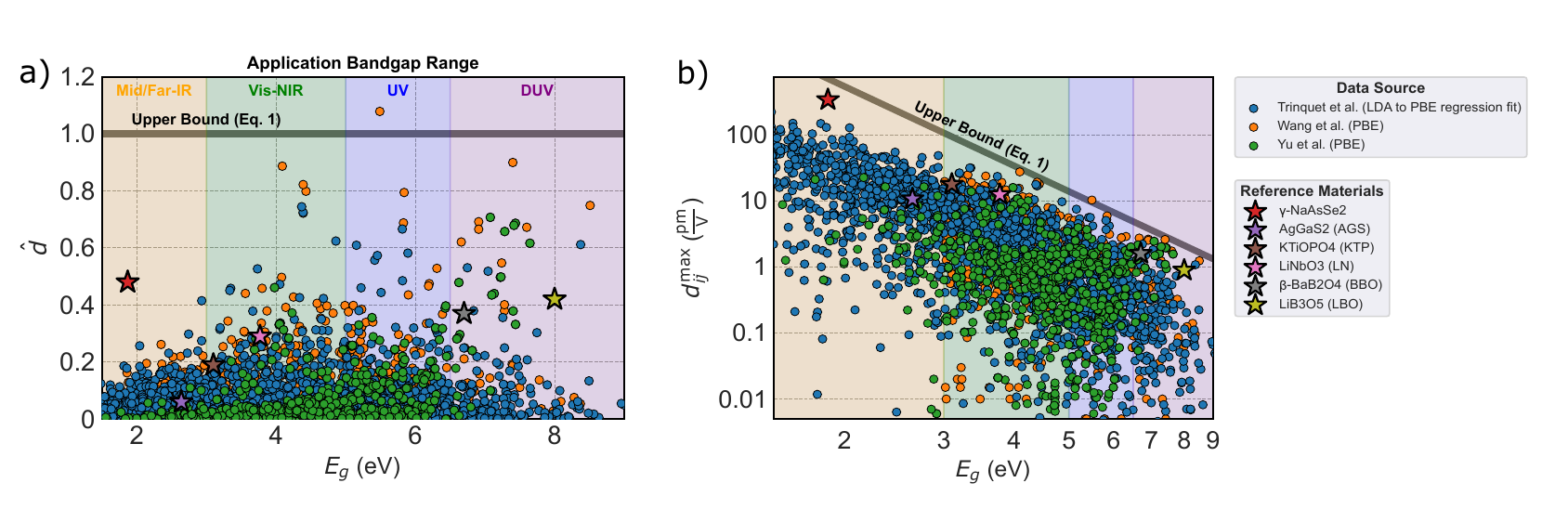}
    \caption{The (a) normalized SHG response $\hat{d}$ and the (b) maximum SHG tensor component $d_{ij}^{\mathrm{max}}$ plotted against the band gap for the PBE dataset, with experimental data for several reference materials included\cite{zhang2020, bera2010}. Background colors represent the approximate band gap ranges for the four application wavelength regimes for NLO materials for SHG\cite{xie2023}. This illustrates the effectiveness of \autoref{eq:d_hat_simplified} in scaling the SHG response to an approximately uniform scale across the range of band gaps.}
    \label{fig:d_hat_distribution}
\end{figure}

Next, we compare $\hat{d}$ to other empirical relations used by the optical and photonic materials communities.
Moss's rule states that the refractive index is approximately related to the band gap by $n^4E_g\approx\mathrm{const}$.\cite{Moss1985} With $n\approx(1+\chi^{(1)})^2$ for optical frequencies, this implies that the refractive index decreases with increasing band gap as $\chi^{(1)}\propto E_g^{-1/2}$.\cite{Moss1985}
This relationship, derived from a semiclassical oscillator model,  reflects how large band gaps constrain the polarizability of a material, but serves primarily to estimate the refractive index rather than define a physical upper limit.
In contrast, the normalized SHG coefficient $\hat{d}$, defined through \autoref{eq:chi2_upper_bound}, quantifies the balance between $\chi^{(2)}$ and $E_g$, providing a physics-based measure of intrinsic performance based on band theory.
Another empirical rule by Miller states that the order of magnitude of the second-order nonlinear susceptibility follows $\chi^{(2)}\propto(\chi^{(1)})^3$.\cite{Miller1964,Ettoumi2010-gu}
Putting these two relationships together yields a predicted relationship of $\chi^{(2)}\propto E_g^{-3/2}$, affirming an inverse relationship although with a weaker scaling law than that predicted from the two-band model.
Alternatively, by using Sellmeier oscillators to represent the dielectric response as a function of frequency, it may be shown that $\chi^{(1)}\propto E_g^{-2}$.
Again using Miller's rule, this results in a predicted second-order scaling relationship of $\chi^{(2)}\propto E_g^{-6}$.
Our approach, in contrast, is conceptually more analogous to Kuzyk's definition of the normalized molecular hyperpolarizability, $\beta_N = \beta / \beta_{\mathrm{max}}$,\cite{kuzyk2006} where $\beta_{\mathrm{max}}$ represents the theoretical maximum hyperpolarizability for a given molecular system. 
While both $\hat{d}$ and $\beta_N$ serve to place second-order NLO  responses in a broader theoretical context, $\hat{d}$ is specifically tailored to solids and relates directly to the SHG coefficient, rather than molecular hyperpolarizability.

To examine the utility of the normalized SHG descriptor, \autoref{fig:d_hat_distribution}a shows the distribution of $\hat{d}$ with $E_g$ for the PBE dataset. 
Experimental data for common, well-known SHG materials is plotted for reference.\cite{zhang2020, bera2010} 
We find that the distribution of $\hat{d}$ remains relatively uniform across a wide range of band gaps.
State-of-the-art high performing materials such as $\gamma$-$\mathrm{Na_2AsSe_2}$ and $\mathrm{LiB_3O_5}$ (LBO) have band gap values that differ by a factor of $\sim$4 and SHG coefficients that differ by over three orders of magnitude.\cite{zhang2020, bera2010}
The $\hat{d}$ values for these two materials, however, are very similar as shown in \autoref{fig:d_hat_distribution}a.
This demonstrates that $\hat{d}$ enables meaningful comparison of NLO performance across  materials with vastly different electronic structures. 
In contrast, direct comparison using $d_{ij}^{\mathrm{max}}$, as in \autoref{fig:d_hat_distribution}b, is complicated by its strong dependence on band gap, varying by several orders of magnitude across the range of relevant band gaps.

Moreover, the definition of $\hat{d}$ in \autoref{eq:d_hat_simplified} provides strong physical intuition and interpretability, allowing us to overcome an important challenge: what constitutes a relatively high SHG response strongly depends on the corresponding band gap since $d_{ij}^{\mathrm{max}}$ may vary by several orders of magnitude. 
Furthermore, $\hat{d}$ quantifies how closely a material approaches the theoretical maximum SHG response. %
Although \autoref{eq:d_hat_simplified} is defined using the maximum component of the SHG tensor, $d_{ij}^{\mathrm{max}}$, the nonlinear optical response of materials is often reported using the Kurtz–Perry effective coefficient, $d_\mathrm{KP}$.\cite{kurtz1968, Clark2022}
Analysis of data from Trinquet \latin{et al.} reveals a Spearman rank correlation coefficient of $\rho_s = 0.99$ between $d_{ij}^{\mathrm{max}}$ and $d_\mathrm{KP}$, indicating a nearly ideal monotonic relationship. 
This strong correlation suggests that substituting $d_\mathrm{KP}$ for $d_{ij}^{\mathrm{max}}$ in the definition of $\hat{d}$ preserves the relative ranking of materials, making it a reliable and practical choice for comparative assessments.
Furthermore, the dimensionless nature and bounded scale (from 0 to approximately 1) of $\hat{d}$ also facilitate straightforward visualization and interpretation.



Importantly, this normalization is also 
advantageous for machine learning workflows. 
$\hat{d}$ is a potential alternative to using $d_{\mathrm{KP}}$ or $d_{ij}^{\mathrm{max}}$ as a screening criterion,\cite{alkabakibi2025, ran2025} as it decouples the contributions to SHG of band gap energies from those of the wavefunction physics.\cite{taghizadeh2021}
Setting a threshold such as $\hat{d}>0.05$ could prove effective for identifying promising materials, especially for higher band gaps where a lower SHG response is expected and a $\chi^{(2)}$ screening threshold is thus less meaningful.
Wang \latin{et al.} previously employed the TTP upper bound,\cite{taghizadeh2021}, as defined in \autoref{eq:chi2_upper_bound}, to introduce the screening metric $F = \chi^{(2)}_{iii} \cdot E_g^4$, which they used to identify 45 promising materials spanning applications from the mid-infrared to deep ultraviolet.\cite{wang2024} 
Since $\hat{d}$ adopts the same scaling relation as $F$, it similarly serves as a useful guide for identifying high-performance materials across a broad range of band gaps, including those often overlooked in conventional searches. 
Unlike $F$, however, the normalized form of $\hat{d}$ offers additional physical insight by expressing the SHG response as a fraction of the quantum-mechanical upper limit. 
A discussion of $\hat{d}$ as a motif indicator is provided in SI.


\subsection{Transferability}
The strong performance of $\hat{d}$ as a descriptor highlights its utility in identifying high-performing materials. 
However, practical screening efforts often rely on optical properties computed using local or semilocal functionals within DFT rather than hybrid functionals, due to their lower computational cost \cite{Rondinelli2015}. 

\begin{figure}[t]
    \includegraphics[width=0.975\linewidth]{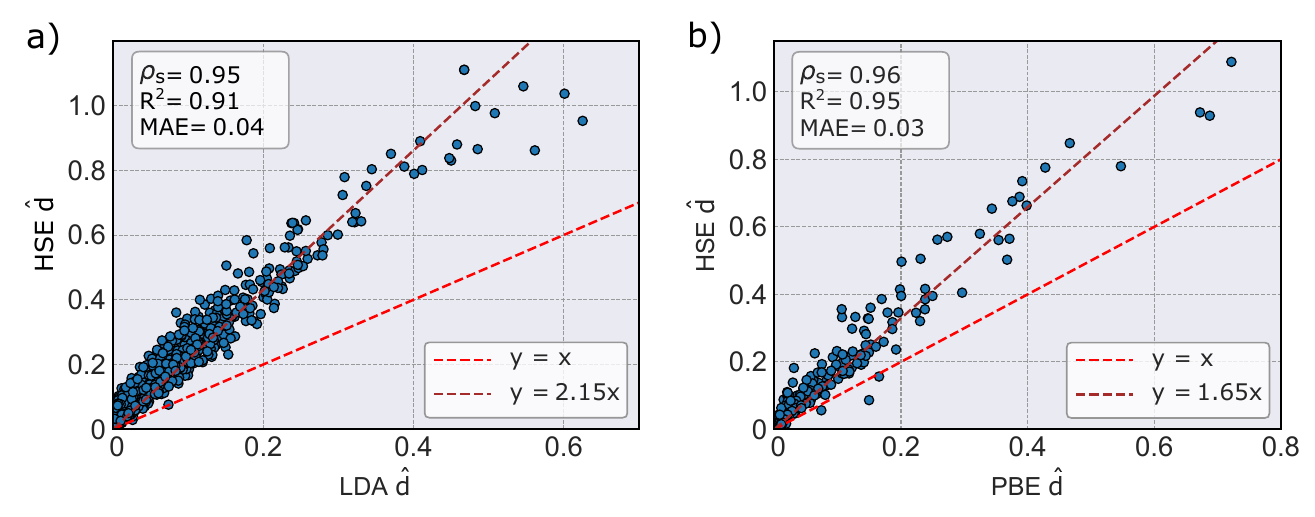}
    \caption{The linear relationships between $\hat{d}$ for (a) HSE and LDA-computed data and between (b) HSE and PBE-computed data, supported by $R^2$, Spearman's $\rho_s$, and MAE values.}
    \label{fig:hse_comparisons}
\end{figure}

To assess the robustness of $\hat{d}$ under these conditions, we examine its behavior across different levels of theory.
Crucially, \autoref{fig:hse_comparisons} shows that $\hat{d}$ values computed using LDA and PBE functionals exhibit strong correspondence with those obtained from HSE,\cite{trinquet2025, lentz2020} enabling reliable mapping between them. 
Indeed, we find $\rho_s$ of 0.95 (0.96) confirm a strong monotonic relationship between HSE and LDA (PBE) derived $\hat{d}$ values, indicating that the relative ranking of materials is preserved across functionals.
Furthermore, linear fits between HSE and LDA (PBE) $\hat{d}$ values yield high coefficients of determination ($R^2$) of 0.95 (0.91), with corresponding slopes of 2.15 (1.65). 

These results implies that screening and ranking based on lower-cost data can yield comparable results, even if the absolute magnitudes differ.
These results reinforce the robustness of $\hat{d}$ as a performance descriptor, demonstrating its broad applicability for evaluating and ranking nonlinear optical materials across different levels of theory and data fidelity.
These results demonstrate that screening and ranking based on lower-cost data can yield comparable outcomes, even if the absolute magnitudes of $\hat{d}$ differ. 
This reinforces the robustness of $\hat{d}$ as a performance descriptor and highlights its broad applicability for evaluating NLO  materials using different levels of data fidelity.

\subsection{Limitations} 
As shown in \autoref{fig:bound}, the scaling law begins to break down at low band gaps—specifically below $\sim3\,\mathrm{eV}$. 
This deviation may be partially attributed to the methodologies used in constructing the underlying databases.
Trinquet \latin{et al.} excluded materials with LDA-calculated $d_{\mathrm{KP}}$ values exceeding 170~$\mathrm{pm\,V}^{-1}$, as well as compounds that were statistical outliers in the ($E_g$, $d_{\mathrm{KP}}$) space.\cite{trinquet2025} 
These filtering criteria may have inadvertently removed low-$E_g$, high-$\chi^{(2)}$ materials, limiting the data available to validate the scaling law in this regime.
Similarly, Wang \latin{et al.} excluded compounds containing transition metals with partially filled 3d orbitals and those exhibiting nonzero magnetization, due to challenges in reliably determining their magnetic ground states.\cite{wang2024} 
Additionally, materials with more than 20 atoms per unit cell were omitted due to limited computational constraints.
Collectively, these screening choices may have led to an under-representation of materials with low band gaps and potentially large $\chi^{(2)}$ values, contributing to the observed breakdown of the scaling law in this region.
The apparent breakdown suggests a potential bias against promising materials in this regime. 
Importantly, however, NLO materials must generally satisfy the condition $E_g > 2\omega$ to suppress two-photon absorption and avoid resonant enhancement of the SHG signal. 
This requirement varies depending on the target application, as illustrated in \autoref{fig:d_hat_distribution}. 
In practice, $E_g$ often needs to be several times larger than $2\omega$ to ensure transparency and minimize losses.
Consequently, $\hat{d}$ is most appropriately applied within application-specific band gap windows. 
For example, materials intended for mid- or far-infrared applications typically require band gaps exceeding 3\,eV, while those for visible, near-infrared, ultraviolet, or deep-ultraviolet applications demand even larger gaps.\cite{qingchen2024} 
Within these ranges, the scaling law remains valid and $\hat{d}$ retains its effectiveness as a comparative metric.
Therefore, although the scaling law may not be validated for low-$E_g$, high-$\chi^{(2)}$ materials, this limitation does not detract from the general utility of $\hat{d}$ in identifying top-performing candidates for technologically relevant NLO applications.
Finally, while $\hat{d}$ serves as a valuable screening metric for nonlinear optical (NLO) performance, it represents only one aspect of material suitability.\cite{Halasyamani_2018} 
Practical deployment of NLO materials also depends on other critical properties, such as birefringence, thermal conductivity, phase stability, and ease of crystal growth. 
These considerations reinforce the role of $\hat{d}$ as a useful initial filter in the discovery process, but emphasize that comprehensive evaluation must incorporate a broader set of criteria.

\section{Conclusion}

The strong band gap-dependence of $\chi^{(2)}$ complicates the quantification of relative SHG performance, which can limit the efficiency of screening promising NLO materials across the entire range of relevant band gaps. 
In this work, we thoroughly validated the scaling relation of the theoretical upper bound on $\chi^{(2)}$ using a combination of new, large NLO datasets, supporting the definition of a normalized SHG coefficient, $\hat{d}$. 
The framing of this dimensionless descriptor as a fraction of the theoretical maximum SHG response enables interpretable comparisons between NLO materials with a wide range of band gaps. 
$\hat{d}$ also serves as an interpretable label for ML models, and opens the door for insightful feature analysis. 

The primary limitation of $\hat{d}$ stems from the ambiguous performance of the upper bound at low band gaps due to dataset biases.
Nonetheless, $\hat{d}$ serves to make screening workflows more efficient, accelerating the discovery of improved NLO materials.
Future efforts could explore the integration of $\hat{d}$ into machine learning frameworks as an alternative to screening based on $E_g$ and $d_{\mathrm{KP}}$ or $d_{ij}^{\mathrm{max}}$ separately. 
As databases of computed NLO properties grow, we expect that ML methods could exploit $\hat{d}$ to enable de novo  materials design insights. 
%

\begin{acknowledgement}
A.\ G.\ J.\ Nyiri thanks the McCormick Advisory Council for funding this project via the MAC Summer Research Grant.
M.\ J.\ W.\ and J.\ M.\ R.\ were supported by the Air Force Office of Scientific Research under Grant No.\ FA9550-23-1-0658.

\end{acknowledgement}

\begin{suppinfo}
The Supporting Information is available free of charge on the ACS Publications website at DOI: 
Data Sources, Top PBE Entries, and Motif Indicator Analysis (PDF).
\end{suppinfo}

\bibliography{references}

\begin{tocentry}
\begin{center}
\includegraphics[width=3.25in]{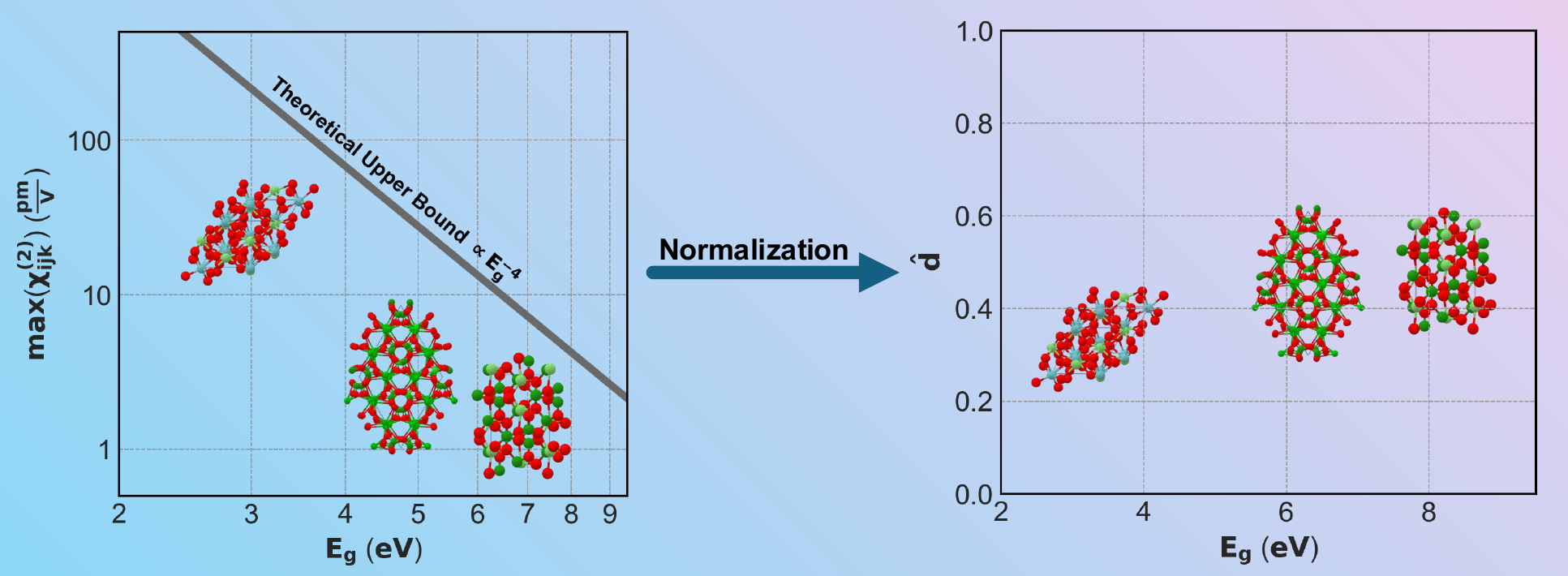}
\end{center}
\end{tocentry}

\end{document}